\documentclass[prl,twocolumn,preprintnumbers,amsmath,amssymb]{revtex4}

\usepackage{epsfig}
\usepackage[colorlinks=true,linkcolor=blue]{hyperref} 
\usepackage{color}
\usepackage{ulem}
\usepackage{dcolumn}
\DeclareGraphicsExtensions{.pdf}

\begin{document}

\title{Simultaneous Determination of Fragmentation Functions and Test on Momentum Sum Rule}

\author{Jun~Gao$^{1,2}$,~ChongYang~Liu$^{1,2}$,~XiaoMin~Shen$^{1,2,3}$,~Hongxi Xing$^{4,5,6}$,~Yuxiang Zhao$^{6,7,8,9}$}

\affiliation{
    $^1$ INPAC, Shanghai Key Laboratory for Particle Physics and Cosmology \\ \& School of Physics and Astronomy, Shanghai Jiao Tong University, Shanghai 200240, China \\
    $^2$Key Laboratory for Particle Astrophysics and Cosmology (MOE), Shanghai 200240, China\\
    $^3$Deutsches Elektronen-Synchrotron DESY, Notkestr. 85, 22607 Hamburg, Germany\\
    $^4${Key Laboratory of Atomic and Subatomic Structure and Quantum Control (MOE), Guangdong Basic Research Center of Excellence for Structure and Fundamental Interactions of Matter, Institute of Quantum Matter, South China Normal University, Guangzhou 510006, China}\\
    $^5${Guangdong-Hong Kong Joint Laboratory of Quantum Matter, Guangdong Provincial Key Laboratory of Nuclear Science, Southern Nuclear Science Computing Center, South China Normal University, Guangzhou 510006, China}\\
    $^6${Southern Center for Nuclear-Science Theory (SCNT), Institute of Modern Physics, Chinese Academy of Sciences, Huizhou 516000, China}\\
    $^7${Institute of Modern Physics, Chinese Academy of Sciences, Lanzhou, Gansu 730000, China}\\
    $^8${University of Chinese Academy of Sciences, Beijing 100049, China}\\
    $^9${Key Laboratory of Quark and Lepton Physics (MOE) and Institute of Particle Physics, Central China Normal University, Wuhan 430079, China}
 }

\email{\\ {jung49@sjtu.edu.cn} \\ {liucy1999@sjtu.edu.cn}\\
{xmshen137@sjtu.edu.cn}\\
{hxing@m.scnu.edu.cn}\\
{yxzhao@impcas.ac.cn}\\}

\begin{abstract}

We perform a simultaneous global analysis of hadron fragmentation functions (FFs) to various charged hadrons ($\pi^{\pm}$, $K^{\pm}$ and $p/\bar{p}$) at next-to-leading order in QCD. 
The world data include results from electron-positron single-inclusive annihilation, semi-inclusive deep inelastic scattering, as well as proton-proton collisions including jet fragmentation measurements for the first time which lead to strong constraints on the gluon fragmentations. 
By carefully selecting hadron kinematics to ensure the validity of QCD factorization and the convergence of perturbative calculations, we achieve a satisfying best fit with $\chi^2/$d.o.f.$=0.90$. 
The total momentum of $u$, $d$ quarks and gluon carried by light charged hadrons have been determined precisely, urging precision determinations of FFs to neutral hadrons for a test of fundamental sum rules in QCD fragmentation.

\end{abstract}
\pacs{}
\maketitle

\pagebreak
\newpage

\begin{figure}[htbp]
  \centering
  \includegraphics[width=0.47\textwidth]{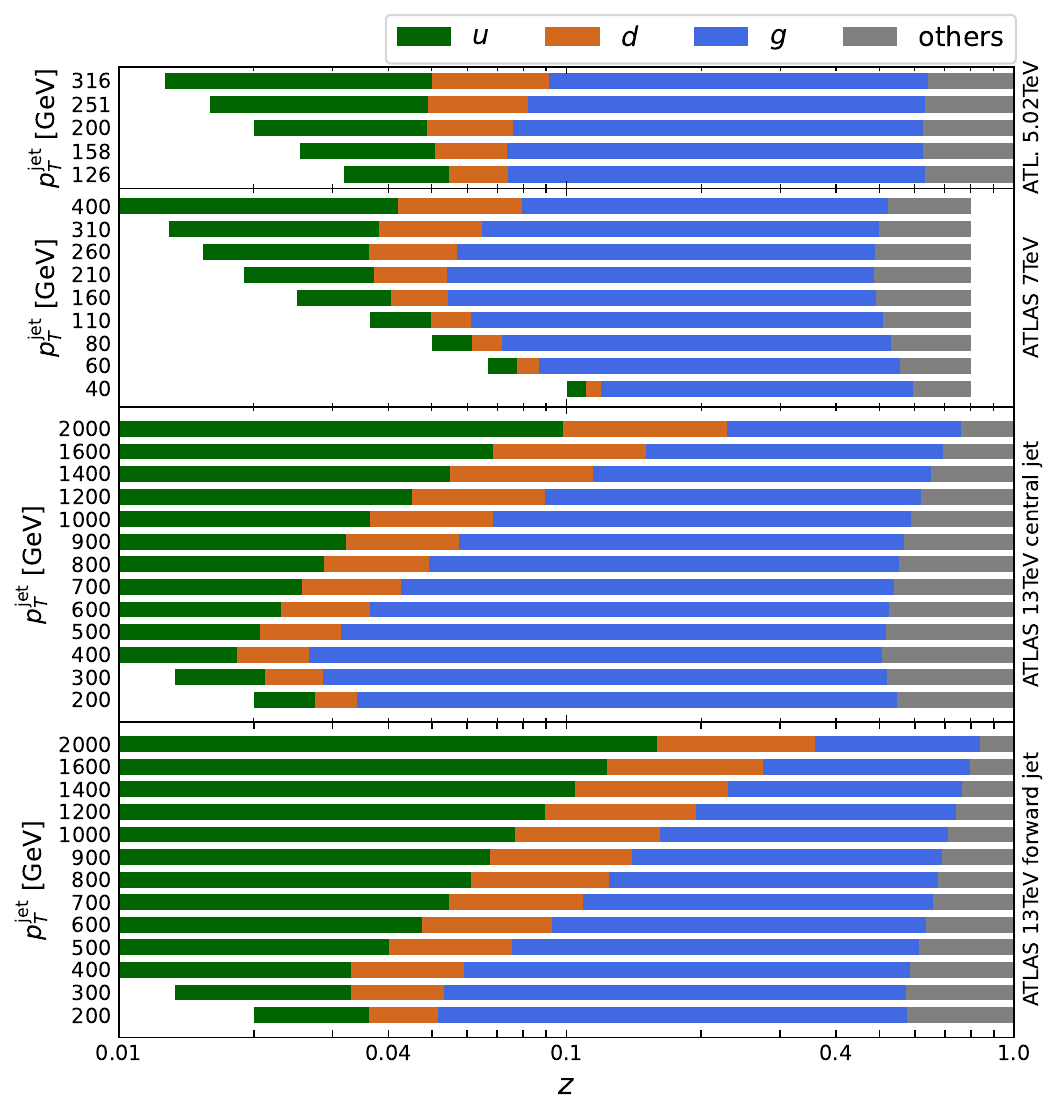}
	\caption{
    Coverage on the momentum fraction $z$ for all subsets of the ATLAS jet fragmentation measurements \cite{ATLAS:2018bvp, ATLAS:2011myc, ATLAS:2019rqw} after the kinematic selections.
	The left end of the bar indicates the lower limit of
    $z$ of the subset and the width of each colored region represents the relative portion of jet flavors, including jets from $u$, $d$ quarks, gluon, and other partons, shown in linear scale.
	}
  \label{Fig:kinejet}
\end{figure}

\noindent{\it Introduction.--}
Single inclusive hadron production has been extensively measured in electron-positron single-inclusive annihilation (SIA), semi-inclusive deep inelastic scattering (SIDIS), as well as proton-proton collisions ($pp$) with the collision energy ranging from $\sqrt{s}=2.23$ GeV at BEPCII \cite{BESIII:2022zit} to $\sqrt{s}=13$ TeV at the Large Hadron Collider \cite{LHCb:2022rky}.
Such processes can be factorized in terms of the hadron fragmentation functions (FFs) within the framework of factorization theorem of quantum chromodynamics (QCD) \cite{Collins:1989gx}. 
The FFs are crucial for comprehending color confinement in QCD, as can be dated back to~\cite{Berman:1971xz,Field:1977fa}.
They not only represent the transition of quarks and gluons into color singlet hadrons, but also play a vital role in accurately probing the internal nucleon structure and the transport properties of partons inside quark-gluon plasma created in heavy ion collisions~\cite{Metz:2016swz,JET:2013cls}. 
Analogous to parton distribution functions (PDFs), the fundamental properties of FFs are rooted in their physical interpretation as number densities and the associated sum rules, with renewed interest regarding the validity of these properties \cite{Collins:2023cuo, Pitonyak:2023gjx}.
Because of their nonperturbative nature and the complexity involved in defining all out states, FFs cannot be computed and their fundamental properties cannot be justified from first principle of QCD. 
Therefore, it is of paramount importance to rigorously determine the FFs through analyses of worldwide data.
Significant efforts have been dedicated to the phenomenological extractions utilizing various data samples, see, e.g., DSS \cite{deFlorian:2007ekg}, HKNS \cite{Hirai:2007cx}, AKK \cite{Albino:2008fy}, SGK \cite{Soleymaninia:2018uiv}, NNFF \cite{Bertone:2018ecm}, MAPFF~\cite{Khalek:2021gxf}, and JAM \cite{Moffat:2021dji}. 
In this Letter, we present novel results on FFs for light charged hadrons, specifically $\pi^{\pm}$, $K^{\pm}$, and $p/\bar{p}$, from a simultaneous determination at next-to-leading order (NLO) in QCD using global datasets of SIA, SIDIS, and $pp$ collisions. 
A stringent selection criterion has been implemented on the kinematics of the fragmentation processes to ensure the validity of leading twist collinear factorization and the associated perturbative calculations of QCD. 
Additionally, we have incorporated residual theory uncertainties into the analysis together with various improvements on implementations of numerical calculations.
To the best of our knowledge, our work represents the first joint global analysis of FFs, and also marks the first inclusion of jet fragmentation measurements, for light charged hadrons. 
The comprehensive analysis provides a state-of-the-art determination of FFs, allowing for a test on the fundamental law of momentum sum rule.
\noindent{\it Theoretical setup and data characteristics.--}
The global analysis of FFs requires a parametrization form at the initial scale $Q_0$. We take the following form for a parton $i$ fragments to a charged hadron $h$,
\begin{equation}\label{eq:para}
z D^{h}_{i}\left(z, Q_{0}\right)=z^{\alpha^h_i}(1-z)^{\beta^h_i} \exp \left(\sum_{n=0}^m a^h_{i,n}(\sqrt {z})^{n}\right),
\end{equation}
which is positively defined, and $\{\alpha, \beta, a_{n}\}$ are free parameters to be determined.
The $Q_0$ is set to be 5 GeV and a zero-mass scheme is utilized for heavy quarks with active quark flavors $n_f=5$.
The isospin or flavor asymmetry is allowed for light quark fragmentation.
For instance, we assume $D^{\pi^+}_{u}$ and $D^{\pi^+}_{\bar d}$ have the same shape but independent normalization, similar for  $D^{K^+}_{ u}$ and $D^{K^+}_{\bar s}$.  
The degree of polynomials $m$ has been increased till no significant improvements of fit are observed, with the final values varying from 0 to 2 depending on the flavors of parton and hadron.
The total number of free parameters is 63 for $\pi^+$, $K^+$, and $p$ together.
The FFs are evolved to higher scales using two-loop timelike splitting kernels~\cite{Stratmann:1996hn} implemented in HOPPET~\cite{Salam:2008qg}, to maintain consistency with the NLO analysis.
Theoretical calculations of the differential cross sections are carried out at NLO in QCD with the FMNLO program~\cite{Liu:2023fsq}, which are accelerated with the interpolation grid and fast convolution algorithms.
For calculations involving initial hadrons, CT14 NLO PDFs~\cite{Dulat:2015mca} are used with $\alpha_S(M_Z)=0.118$~\cite{Zyla:2020zbs}. 
The Impacts of using alternative PDFs are small in general (see Supplemental Material \cite{SM}.)
The central values of the renormalization and fragmentation scales ($\mu_{R,0}$ and $\mu_{D,0}$) are set to the momentum transfer $Q$ for both SIA and SIDIS.
In the case of $pp$ collisions, the central values of the factorization scale ($\mu_{F,0}$) and renormalization scale are set to half of the sum of the transverse mass of all final state particles.
The fragmentation scale is set to the transverse momentum of the leading parton for inclusive hadron production, and to the transverse momentum of the jet multiplied by the jet cone size for fragmentation inside jet~\cite{Kaufmann:2015hma}.
Theoretical uncertainties are included in the covariance matrix of $\chi^2$ calculations, and are assumed to be fully correlated among points in each subset of the data.
These uncertainties are estimated by the half width of the envelope of theoretical predictions of the 9 scale combinations of $\mu_F/\mu_{F,0}=\mu_R/\mu_{R,0}=\{1/2,1,2\}$ and $\mu_D/\mu_{D,0}=\{1/2,1,2\}$. 
We note FFs to pions at next-to-next-to-leading order (NNLO) are available in~\cite{Borsa:2022vvp,AbdulKhalek:2022laj} based on analyses of SIA and SIDIS data, though gluon FFs are less constrained due to absence of data from $pp$ collisions.
An extension of our analysis to NNLO is ongoing,  given the available perturbative ingredients.  
We consider datasets from SIA, SIDIS, and $pp$ collisions as listed in Table~\ref{Tab:chi2}.
In particular, the measurements on unidentified and identified charged hadron production from fragmentation inside inclusive and $Z/\gamma$ tagged jets by CMS, ATLAS and LHCb~\cite{ATLAS:2019dsv, CMS:2018mqn, ATLAS:2020wmg, CMS:2021otx, ATLAS:2018bvp, ATLAS:2011myc, ATLAS:2019rqw, LHCb:2022rky} are incorporated for the first time.
Additionally, we include inclusive hadron production measurements from ALICE~\cite{ALICE:2014juv, ALICE:2019hno,ALICE:2020jsh} and STAR~\cite{STAR:2011iap}, and consider only ratios of inclusive cross sections of different charged hadrons or of different collision energies to minimize the impact of normalization uncertainties.
For SIA, we incorporate a comprehensive set of data from TASSO, TPC below the $Z$ pole~\cite{TASSO:1988jma,TPCTwoGamma:1988yjh}, from OPAL, ALEPH, DELPHI, and SLD at the $Z$ pole~\cite{OPAL:1994zan,ALEPH:1994cbg,DELPHI:1998cgx,SLD:2003ogn}, and from OPAL and DELPHI above the $Z$ pole~\cite{OPAL:2002isf,DELPHI:2000ahn}.
For SIDIS, we utilize data on the total rate and charge asymmetry of production of unidentified charged hadrons from H1 and ZEUS at high $Q^2$~\cite{H1:2007ghd,H1:2009lef, ZEUS:2010mrq}. 
They are accompanied by measurements on production of identified charged hadrons from COMPASS at relatively low $Q^2$ with isoscalar (06$I$) or proton (16$p$) targets~\cite{COMPASS:2016xvm,COMPASS:2016crr,Pierre:2019nry}, only for the two subsets with largest Bjorken-$x$ and the highest inelasticity.
The impact of inclusion of COMPASS datasets with even lower $Q^2$ values are also studied (see Supplemental Material \cite{SM}).
In our analysis it is assumed that the measured cross sections on unidentified charged hadrons are a combination of charged pion, kaon, and proton, while the residual contribution is negligible.
Strict selection criteria are applied to the kinematics of data points.
Specifically, we exclusively select data points corresponding to momentum fractions $z>0.01$ at leading order except for single inclusive hadron production in $pp$ collisions. 
Additionally, it is required that $p_{T,h}(E_h)>$ 4~GeV for data from $pp$ collisions (SIA and SIDIS) except for COMPASS where the $Q$ value can be as low as 3.7~GeV, with the hadron energy being measured in the Breit frame for SIDIS.
All the aforementioned data can be categorized into 138 subsets based on the range of jet $p_T$ for jet fragmentation, $Q^2$ for SIDIS, and collision energy for inclusive hadron production at $pp$ collisions and SIA.
As an illustrative example, we present the $z$ coverage for all subsets of the ATLAS jet fragmentation measurements after the kinematic selections in Fig.~\ref{Fig:kinejet}, together with flavor decomposition of the jet.
They include different $p_T$ bins from the inclusive jet production at 7 and 5.02 TeV, as well as from both the central and forward jet in dijet production at 13 TeV.
It is clearly shown that the gluon fragmentation is dominant for jet production at low-$p_T$, e.g., 200~GeV, and at small rapidity.
This highlights the strong constraints from jet fragmentation on the gluon FFs. 

\noindent{\it The fit.--}
The log-likelihood functions $\chi^2$ are calculated for each subset using predictions from the prescribed theory and the covariance matrices constructed from both experimental and theoretical uncertainties.
A best fit for the parameters of the FFs is determined by minimizing the total $\chi^2$ using the MINUIT program~\cite{James:1975dr}.
In Table.~\ref{Tab:chi2}, we provide a summary of various results demonstrating the quality of the best fit.
The global $\chi^2$ is 1231.5 units for a total of 1370 data points, indicating overall good agreement between theory and data.
The $\chi^2/N_{pt}$ values are all below 1 for groups of data of $pp$ collisions, SIA, and SIDIS.

\begin{table}[]
  \centering
\begin{tabular}{|c|c|c|c|}
\hline
Experiments & \textbf{$N_{pt}$} & \textbf{$\chi^2$} & \textbf{$\chi^2/N_{pt}$} \\ \hline
ATLAS jets $^\dag$             &446	&350.8	&0.79\\ 
ATLAS $Z/\gamma$+jet $^\dag$   &15	&31.8	&2.12\\
CMS $Z/\gamma$+jet $^\dag$   &15	&17.3	&1.15\\ 
LHCb $Z$+jet          &20	&30.6	&1.53\\
ALICE inc. hadron      &147	&150.6	&1.02\\ 
STAR inc. hadron      &60	&42.2	&0.70\\ \hline
$pp$ sum             &703	&623.3	&0.89\\ \hline
TASSO                   &8	&7.0	&0.88\\
TPC                    &12	&11.6	&0.97\\
OPAL                  &20	&16.3	&0.81\\
OPAL (202 GeV) $^\dag$       &17	&24.2	&1.42\\ 
ALEPH                &42	&31.4	&0.75\\ 
DELPHI             &78	&36.4	&0.47\\   
DELPHI (189 GeV)    &9	&15.3	&1.70\\  
SLD                 &198	&211.6	&1.07\\   \hline 
SIA sum             &384	&353.8	&0.92\\    \hline
H1 $^\dag$              &16	&12.5	&0.78\\      
H1 (asy.) $^\dag$        &14	&12.2	&0.87\\     
ZEUS $^\dag$                &32	&65.5	&2.05\\
COMPASS (06$I$)             &124	&107.3	&0.87\\ 
COMPASS (16$p$)          &97	&56.8	&0.59\\ \hline
SIDIS sum             &283	&254.4	&0.90\\ \hline
Global total&1370	&1231.5 &	0.90\\  \hline
\end{tabular}
	\caption{
	The number of data points, $\chi^2$, and $\chi^2/N_{pt}$ for the global datasets, groups of data from $pp$ collisions, from SIA, and from SIDIS, and individual experiments.
    Datasets for production of unidentified charged hadrons are marked with a dagger. 
	}
  \label{Tab:chi2}
\end{table}

\begin{figure}[htbp]
  \centering
  \includegraphics[width=0.47\textwidth]{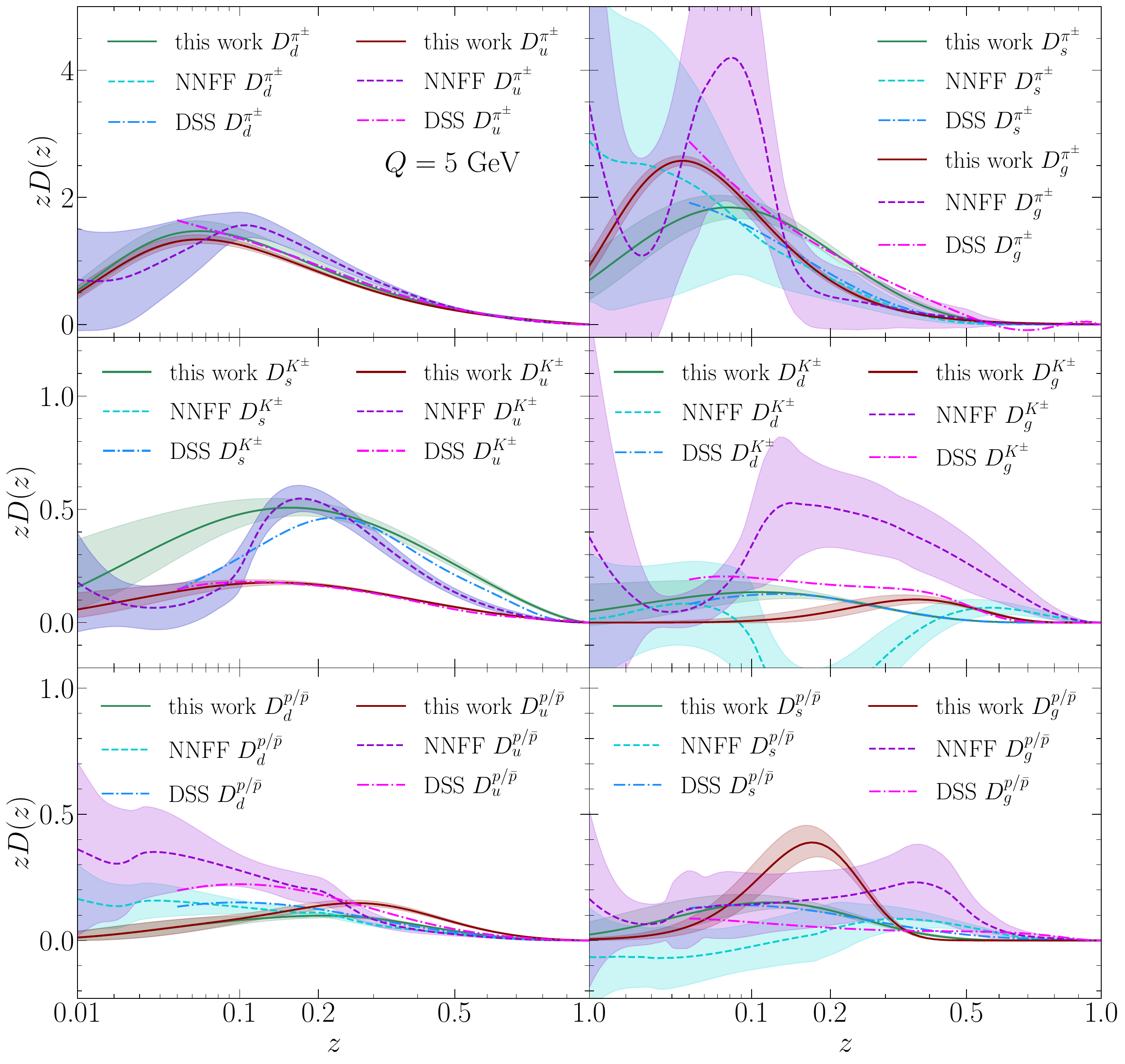}
	\caption{
	Comparison of our NLO fragmentation functions to those from NNFFs and DSS at $Q=5$~GeV.
	The estimated uncertainties of FFs are also shown for NNFFs and for our fit.
	For $\pi^{\pm}$ and $p/\bar p$ the left (right) panel shows results for $u$ and $d$ ($s$ and $g$).
	For $K^{\pm}$ the left (right) panel shows results for $u$ and $s$ ($d$ and $g$).
	}
  \label{Fig:ffs}
\end{figure}

To assess the agreement for each of the 138 subsets, we utilize an effective Gaussian variable
\begin{equation}
    S_E=\frac{(18N_{pt})^{3/2}}{18N_{pt}+1}\left\{{6\over 6-\ln(\chi^2/N_{pt})}-{9N_{pt}-1 \over  9N_{pt} }\right\},
\end{equation}
which ideally follows a normal distribution if $N_{pt}$ is not too small~\cite{ct18}.
The majority of the subsets (132 out of 138) have $S_E$ values smaller than 2, indicating good agreement. 
The histogram of $S_E$ for all subsets in our best-fit closely resembles a Gaussian distribution, but with a mean of -0.33 and a standard deviation of 1.43.
The deviation from the standard normal distribution suggests a potential underestimation of experimental or theoretical uncertainties by an average factor of 1.43.
This motivates a choice of tolerance of $\Delta\chi^2=1.43^2\approx 2$ in our estimation of uncertainties of the FFs using the Hessian method~\cite{ct18}.
\noindent{\it Resulting FFs.--}
Our newly obtained NLO FFs \footnote{The newly obtained FFs are denoted as NPC23, and are publicly available in the form of LHAPDF6 grids on \url{https://fmnlo.sjtu.edu.cn/\~fmnlo/data/NPC23_CH.tar.gz} or from the LHAPDF website.} are compared to those from NNFF1.0~\cite{Bertone:2017tyb} and DSS \cite{Borsa:2021ran, deFlorian:2017lwf, deFlorian:2007ekg} for $u$, $d$, $s$ quark, and gluon at $Q=5$~GeV in Fig.~\ref{Fig:ffs}.
For simplicity, we only show the FFs 
summed over charges for pions, kaons and protons from top to bottom insets.
Reasonable agreement can be observed between our results and DSS for FFs of $u$ and $d$ quarks to $\pi^{\pm}$, and of $u$ quark to $K^{\pm}$.
However, large discrepancies are found for FFs to protons and for FFs of gluon to all three charged hadrons.
Our results show an uncertainty of 3\%, 4\%, and 8\% for FFs of gluon to $\pi^{\pm}$ at $z=0.05,\,0.1$, and $0.3$, respectively, which are significantly improved compared with NNFFs.
The uncertainties are about 4\%, 4\%, and 7\% for FFs of $u$ quark to $\pi^{\pm}$, $K^{\pm}$, and $p/\bar p$ at $z=0.3$.
The high precision of gluon FFs is mostly due to the data of jet fragmentation at the LHC.
Furthermore, the newly included data on proton production from SIDIS and $pp$ collisions lead to better flavor separation and thus the differences observed for FFs to protons.

\begin{figure}[htbp]
  \centering
  \includegraphics[width=0.47\textwidth]{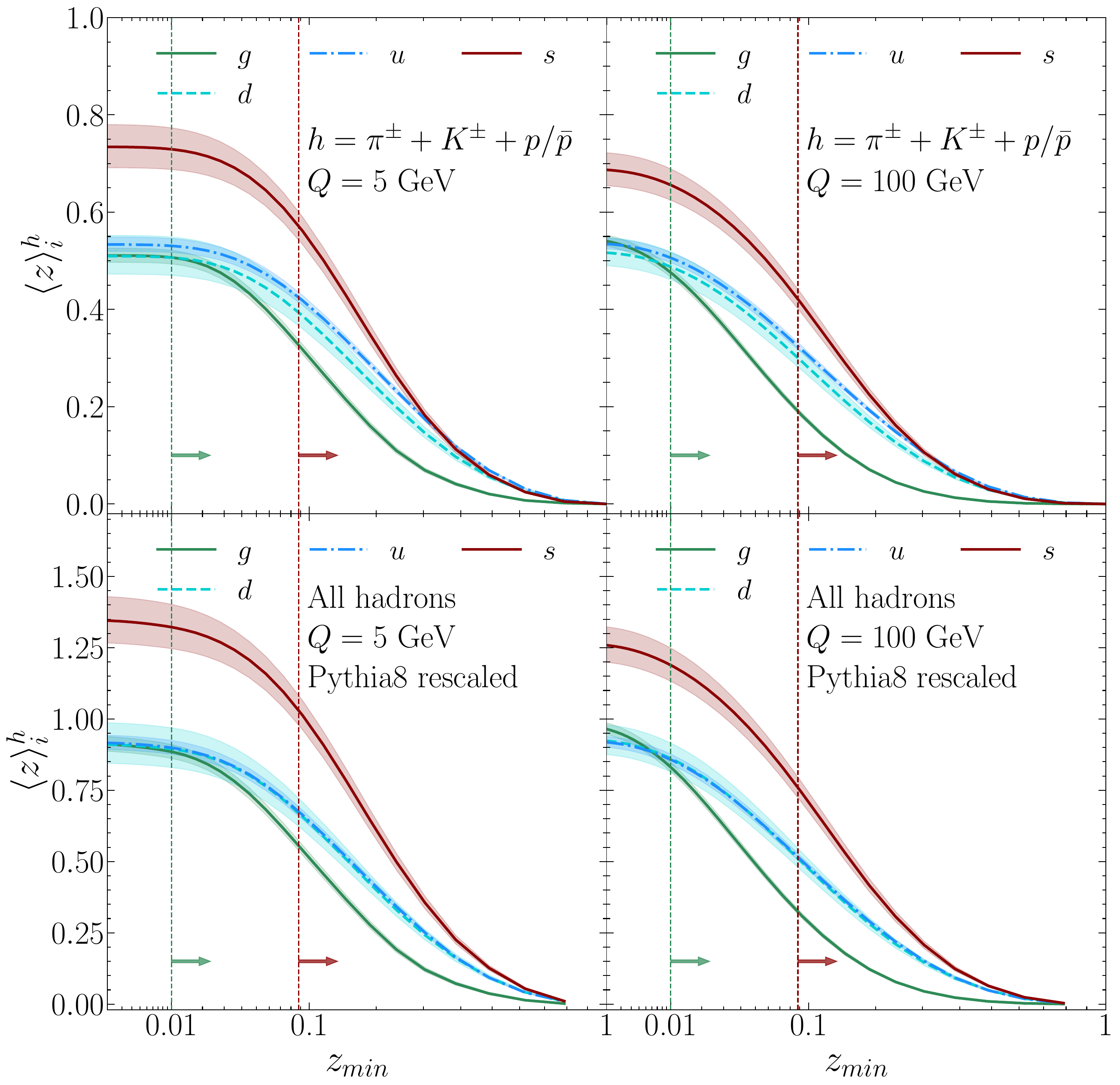}
	\caption{
Total momentum of the partons, including $g$, $u$, $d$, and $s$  quarks, carried by light charged hadrons and all hadrons in the fragmentations, as functions of $z_{\rm min}$.
See the text for details.}
  \label{Fig:mom}
\end{figure}

Fundamental sum rules of FFs arise from their number density interpretation and have been pointed out to be problematic by Rogers and Collins in Ref.\cite{Collins:2023cuo}. 
It is of critical importance to check from a data-driven analysis whether these fundamental properties of FFs are valid. 
That is especially the case for the momentum sum rule of FFs,
\begin{equation}
\sum_h\int_0^1 dz zD_i^h(z,Q) = 1,
\end{equation}
due to the suppression of small-$z$ contributions.
The above momentum sum rule, summed over all hadrons, was approved for each specific flavor of quark and gluon.
We first calculate the total momentum fraction $\langle z\rangle_i^h =\int_{z_{{\rm min}}}^{1}dz zD_i^h(z,Q)$ of parton $i$ carried by hadron $h$ for $z_{\rm min}<z<1$ at initial scale $Q=5$ GeV, where the choice of the lower limit $z_{\rm min}$ is $0.01$ for $g$, $u$, and $d$ quarks, and $0.088$ for $s$, $c$, and $b$ quarks, based on the kinematic coverage of relevant data.
The final results of $\langle z\rangle_i^h$ for light quarks and gluon including Hessian uncertainties are shown in Table.~\ref{Tab:mom}. 
It shows that the three charged hadrons carry approximately 53\% to 50\% of the momentum of $u$, $d$ quarks and gluon.
Our analysis reveals a preference for larger FFs of $s$ quark to $\pi^{\pm}$, with each carrying about 16\% of the total momentum of $s$ quark.
One possible reason is because the SIA measurements on spectrum of $\pi^{\pm}$ also include feed-down contributions from short-lived
strange hadrons~\cite{SLD:2003ogn}.
The total momentum carried by different sets of hadrons as functions of $z_{\rm min}$ for $Q$=5 (left) and 100~GeV (right) are shown in Fig.~\ref{Fig:mom}. 
The vertical dashed lines show the lower limit of $z$ as constrained by data for $g,~u,~d$, and $s$ quark fragmentation, respectively. 
One can see from the top-left figure that $\langle z \rangle_{i=g,u,d}^{h}$ for light charged hadrons reach to a saturation region within the current experimental coverage in contrast to the strange quark.
A reliable test of momentum sum rule also requires FFs to neutral hadrons determined at similar precision which are not yet available.
Alternatively, one can calculate ratios of energies carried by all hadrons and by light charged hadrons as functions of $z_{\rm min}$ using PYTHIA8~\cite{Bierlich:2022pfr} simulations of 
$q\bar q$ and $gg$ production in $e^+e^-$ collisions at 200 GeV.
We can estimate total momentum of FFs to all hadrons by applying the scaling factors derived from PYTHIA8, as shown in the lower panel of Fig.~\ref{Fig:mom}. 
The central values are slightly lower than 1 for $u$, $d$ quarks and gluon when extrapolated into small-$z_{\rm min}$ region.
That is consistent with the momentum sum rule considering the shown uncertainty and additional uncertainties from the scaling factors.
For strange quarks the values can be well above 1 due to both the ambiguity mentioned earlier and the limited coverage of data.
We have tested with more flexible parametrization forms of strange quark FFs or with scaling factors derived at lower energies or from PYTHIA6~\cite{Sjostrand:2006za}, and found the changes of momentum sum for strange quark are small comparing to the shown Hessian uncertainties.  
We leave detailed investigations of this anomaly for a future publication.

\begin{table}[]
  \centering
\begin{tabular}{|c|c|c|c|c|}
\hline
 $\langle z\rangle_i^h$ & $g(z>0.01)$  & $u(z>0.01)$ & $d(z>0.01)$ & $s(z>0.088)$    \\ \hline
\textbf{$\pi^+$} 
 &$0.200	^{+0.008}_{-0.008}$&$0.262^{+0.017}_{-0.016}$&$0.128^{+0.020	}_{-0.019}$&$0.161^{+0.013}_{-0.013}$\\
\textbf{$K^+$} 
 &$0.018	^{+0.004}_{-0.003}$&$0.058^{+0.005}_{-0.004}$&$0.019^{+0.004	}_{-0.004}$&$0.015^{+0.002}_{-0.002}$\\
\textbf{$p$} 
 &$0.035	^{+0.006}_{-0.005}$&$0.044^{+0.004}_{-0.004}$&$0.022^{+0.002	}_{-0.002}$&$0.015^{+0.002}_{-0.002}$\\ \hline
\textbf{$\pi^-$}  
 &$0.200	^{+0.008}_{-0.008}$&$0.128^{+0.020 }_{-0.019}$&$0.299^{+0.054	}_{-0.049	}$&$0.161^{+0.013}_{-0.013}$\\
\textbf{$K^-$} 
 &$0.018	^{+0.004}_{-0.003}$&$0.019^{+0.004}_{-0.004}$&$0.019^{+0.004	}_{-0.004}$&$0.205^{+0.014	}_{-0.013}$\\
\textbf{$\bar p$} 
 &$0.035	^{+0.006}_{-0.005}$&$0.019^{+0.003}_{-0.003}$&$0.019^{+0.003	}_{-0.003}$&$0.015^{+0.002}_{-0.002}$\\ \hline
Sum 
 &$0.507	^{+0.014}_{-0.013}$&$0.531^{+0.015}_{-0.013}$&$0.506^{+0.042	}_{-0.037}$&$0.572^{+0.029}_{-0.028}$\\ \hline
\end{tabular}
	\caption{
	Total momentum of the partons, including $g$, $u$, $d$, and $s$ quarks, carried by various charged hadrons in the fragmentations.
	}
  \label{Tab:mom}
\end{table}

\noindent{\it Conclusions.--}
In summary, we present a joint determination of FFs for charged hadrons from a global analysis at NLO in QCD, which are essential for programs at the upcoming electron-ion colliders~\cite{Accardi:2012qut,Anderle:2021wcy}.
Our analysis demonstrates good agreement between our best-fit predictions and various measurements in SIA, SIDIS and $pp$ collisions. 
Notably, we have included measurements on jet fragmentation at the LHC into the global analysis, resulting in strong constraints on the gluon FFs.
Comparing our results with previous determinations, we find significant differences, in the fragmentation to protons
and also for FFs of nonconstituent quarks and gluon to charged pions.
Additionally, we provide results on the total momentum of partons carried by various charged hadrons, and find that they are much larger for strange quarks than for up and down quarks, which is potentially inconsistent with the momentum sum rule.
%

\quad \\
\noindent \textbf{Acknowledgments.} 
The work of J.G. is supported by the National Natural Science Foundation of China (NSFC) under Grant No.~12275173. H.X. is supported by the NSFC under Grants No.~12022512, No. 12035007, by the Guangdong Major Project of Basic and Applied Basic Research No.~2020B0301030008, No. 2022A1515010683. Y.Z. is supported by the NSFC under Grant No.~U2032105. X.S. is supported by the Helmholtz-OCPC Postdoctoral Exchange Program under Grant No.~ZD2022004.
%



\newpage
\widetext
\clearpage
\begin{center}


\textbf{Supplemental Material for ``Simultaneous Determination of Fragmentation Functions and Test on Momentum Sum Rule"}

\end{center}

\subsection{Alternative fits using different PDF sets}
\begin{figure}[hbp]
  \centering
  \includegraphics[width=0.8\textwidth]{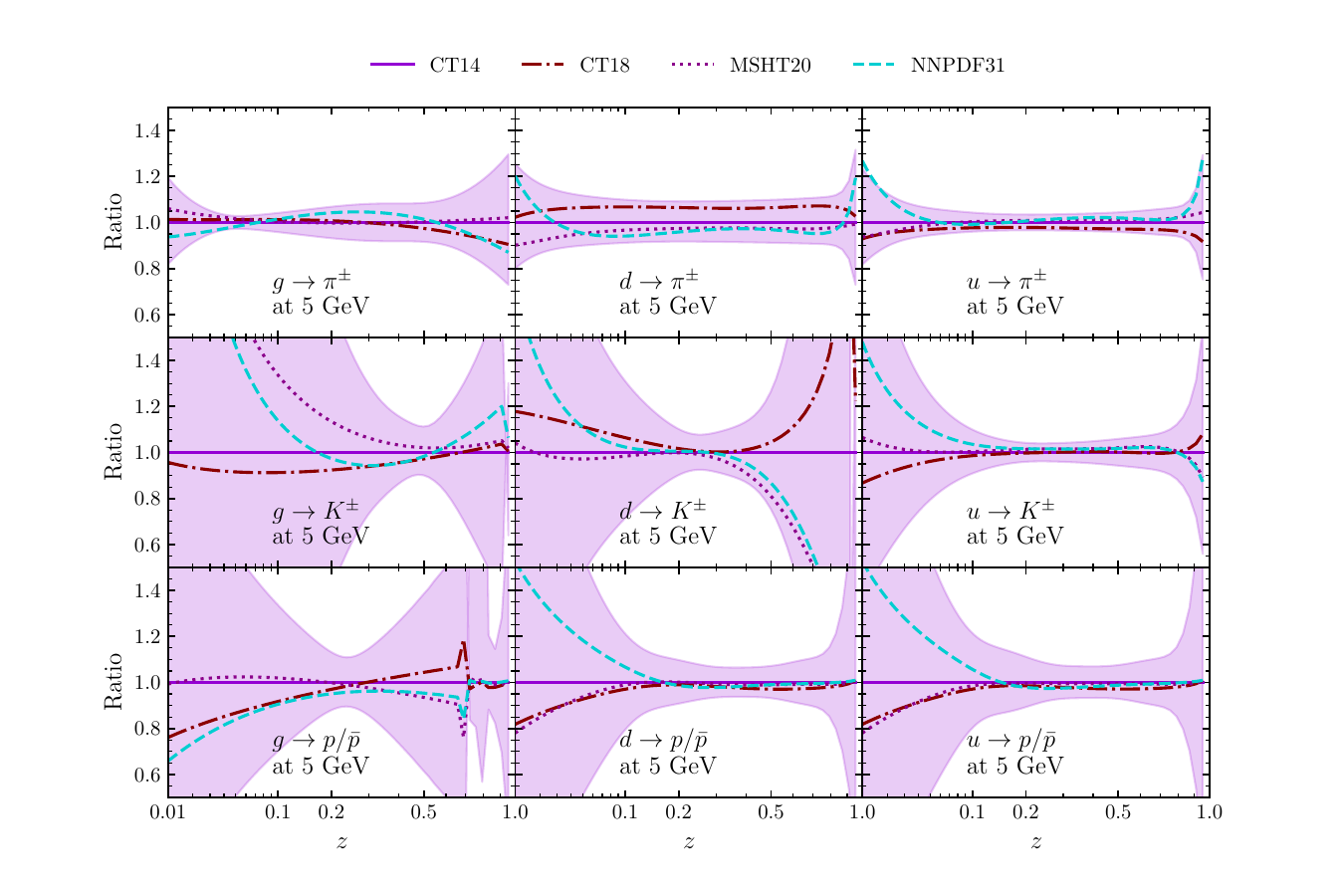}
	\caption{
	Comparisons of FFs from analyses using different NLO PDFs, including CT14, CT18, MSHT20 and NNPDF3.1, for FFs to $\pi^{\pm}$, $K^{\pm}$, and $p/\bar p$ from top to bottom panels.
	From left to right the results correspond to fragmentation of gluon, $d$, and $u$ quarks, respectively.
	All results are normalized to the best-fit FFs using CT14 and the colored bands represent the Hessian uncertainties of FFs.
	}
  \label{Fig:pdfvar}
\end{figure}

We perform alternative fits of the fragmentation functions (FFs) while using CT18~\cite{Hou:2019efy}, MSHT20~\cite{Bailey:2020ooq}, and NNPDF3.1~\cite{NNPDF:2017mvq} parton distribution functions (PDFs) at next-to-leading order (NLO) for theory predictions of data from SIDIS and $pp$ collisions.
The resulted FFs are compared to those from our nominal fit using CT14 NLO PDFs~\cite{Dulat:2015mca} by normalized to the best-fit results, for gluon, $d$ and $u$ quarks to $\pi^{\pm}$, $K^{\pm}$, and $p/\bar p$ respectively, shown in Fig.~\ref{Fig:pdfvar}.
The colored bands represent the Hessian uncertainties of our nominal fit.
In all cases the variations from using different PDFs are within the uncertainty bands.
Dependence on PDFs are minimal especially in kinematic region well constrained by data.
That can be understood because dependence on PDFs are largely cancelled in the numerator and denominator for predictions of all $pp$ data used in this analysis.

\section{Fits including COMPASS low $Q^2$ data}
\begin{figure}[htbp]
  \centering
  \includegraphics[width=0.8\textwidth]{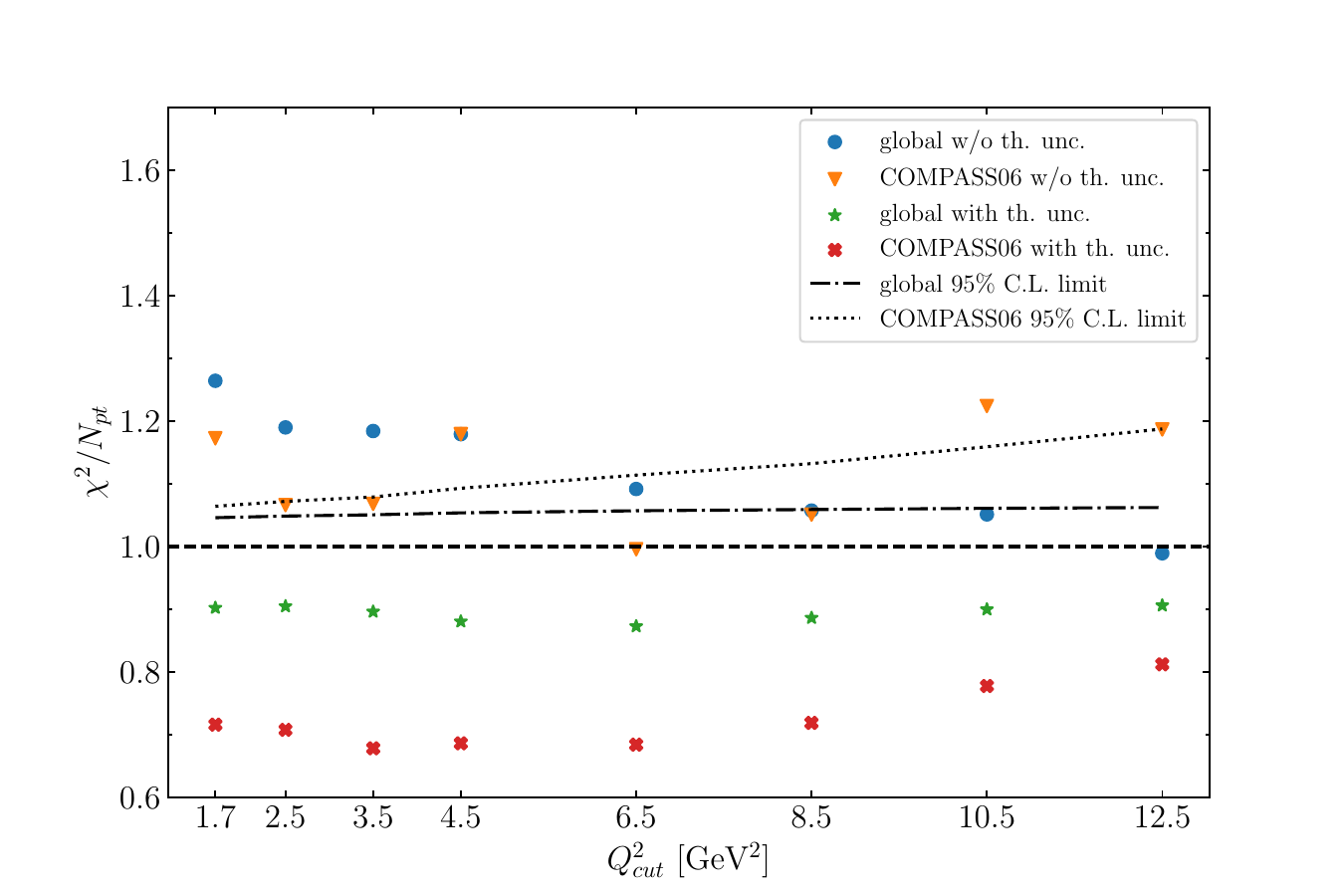}
	\caption{%
 The $\chi^2/N_{\rm pt}$ for the global data or the COMPASS06 data only in the global analysis, as functions of the kinematic cut $Q^2_{\rm cut}$, from fits with and without inclusions of theoretical uncertainties for NLO predictions of COMPASS06 data.
 The dash-dotted and dotted lines correspond to the 95\% C.L. upper limit of $\chi^2$ for the global data and COMPASS06 data respectively.
	}
  \label{Fig:lowq2}
\end{figure}

The COMPASS experiment has measured charged hadron multiplicities in kinematic bins of Bjorken-$x$ from 0.004 to 0.4 and inelasticity $y$ from 0.1 to 0.7, which correspond to a wide range of 0.36 to 60 ${\rm GeV}^2$ on the square of momentum transfer $Q^2$~\cite{COMPASS:2016xvm,COMPASS:2016crr}.   
In our nominal fit we only include the two kinematic bins with $x$ in $[0.14,\,0.18]$ and $[0.18,\,0.4]$, and $y$ in $[0.3,\,0.5]$.
We extend our analysis by including all possible kinematic bins of the COMPASS06 measurement but with a kinematic cut of $Q^2>Q^2_{\rm cut}$.
We have investigated fit quality of the global analysis by varying $Q^2_{\rm cut}$ from 12.5 to 1.7 ${\rm GeV}^2$, corresponding to total number of points for the global (COMPASS06) data from 1400 (154) to 2562 (1316).
The $\chi^2/N_{\rm pt}$ are summarized in Fig.~\ref{Fig:lowq2} for the global $\chi^2$ or the $\chi^2$ of the COMPASS06 data only in the global analysis.
The dash-dotted and dotted lines correspond to the 95\% C.L. upper limit of $\chi^2$ for the global data and COMPASS06 data respectively.
We have tested two different scenarios by including theoretical uncertainties of NLO predictions for COMPASS06 data, same as in our nominal fit, or not including theoretical uncertainties for COMPASS06 data.
In the case of with theoretical uncertainties, the $\chi^2/N_{\rm pt}$ are well below 1 for both the global data and the COMPASS06 data, and for all $Q^2_{\rm cut}$ shown.
The $\chi^2$ increase significantly once excluding theoretical uncertainties for COMPASS06 predictions.
The $\chi^2/N_{\rm pt}$ of COMPASS06 data are stabilized close to the upper limits except for $Q^2_{\rm cut}=1.7$ ${\rm GeV^2}$.
However, the $\chi^2/N_{\rm pt}$ of global data increases steadily with the lowering of $Q^2_{\rm cut}$, and is far above the upper limit for $Q^2_{\rm cut}=4.5$ ${\rm GeV^2}$ or lower. 
The observed trend is similar to that shown in Table.1 of Ref.~\cite{Borsa:2022vvp} but the deterioration of $\chi^2$ happened at a larger $Q^2_{\rm cut}$ in our analyses, possibly due to much more high-$Q^2$ data included in our analyses. 

\end{document}